\documentstyle[12pt,epsf]{article}
\begin{document}

\begin{titlepage}

\vspace{2cm}

\begin{center}


{\LARGE Measurement of the Associated $\gamma + \mu^\pm$ Production 
Cross Section in $p \bar p$ Collisions at $\sqrt{s} = 1.8$~TeV }

\vspace{0.75cm}


\font\eightit=cmti8
\def\r#1{\ignorespaces $^{#1}$}
\hfilneg
\begin{sloppypar}
\noindent
F.~Abe,\r {13} M.~G.~Albrow,\r 7 S.~R.~Amendolia,\r {22} D.~Amidei,\r {16} 
J.~Antos,\r {28} C.~Anway-Wiese,\r 4 G.~Apollinari,\r {26} H.~Areti,\r 7 
M.~Atac,\r 7 P.~Auchincloss,\r {25}
% ...
\end{sloppypar}

\vskip .025in
\begin{center}
(CDF Collaboration)
\end{center}

\vskip .025in
\begin{center}
\r 1  {\eightit Argonne National Laboratory, Argonne, Illinois 60439} \\
\r 2  {\eightit Istituto Nazionale di Fisica Nucleare, University of Bologna,
I-40126 Bologna, Italy} \\
\r 3  {\eightit Brandeis University, Waltham, Massachusetts 02254} \\
\r 4  {\eightit University of California at Los Angeles, Los 
Angeles, California  90024} \\  
\r 5  {\eightit University of Chicago, Chicago, Illinois 60637} \\
\r 6  {\eightit Duke University, Durham, North Carolina  27708} \\
\r 7  {\eightit Fermi National Accelerator Laboratory, Batavia, Illinois 
      60510} \\
\r 8  {\eightit Laboratori Nazionali di Frascati, Istituto Nazionale di Fisica
               Nucleare, I-00044 Frascati, Italy} \\
\r 9  {\eightit Harvard University, Cambridge, Massachusetts 02138} \\
\r {10} {\eightit University of Illinois, Urbana, Illinois 61801} \\
\r {11} {\eightit Institute of Particle Physics, McGill University, Montreal 
H3A 2T8, and University of Toronto,\\ Toronto M5S 1A7, Canada} \\
\r {12} {\eightit The Johns Hopkins University, Baltimore, Maryland 21218} \\
\r {13} {\eightit National Laboratory for High Energy Physics (KEK), Tsukuba, 
        Ibaraki 305, Japan} \\
\r {14} {\eightit Lawrence Berkeley Laboratory, Berkeley, California 94720} \\
\r {15} {\eightit Massachusetts Institute of Technology, Cambridge,
        Massachusetts  02139} \\   
\r {16} {\eightit University of Michigan, Ann Arbor, Michigan 48109} \\
\r {17} {\eightit Michigan State University, East Lansing, Michigan  48824} \\
\r {18} {\eightit University of New Mexico, Albuquerque, New Mexico 87131} \\
\r {19} {\eightit Osaka City University, Osaka 588, Japan} \\
\r {20} {\eightit Universita di Padova, Istituto Nazionale di Fisica 
        Nucleare, Sezione di Padova, I-35131 Padova, Italy} \\
\r {21} {\eightit University of Pennsylvania, Philadelphia, 
        Pennsylvania 19104} \\   
\r {22} {\eightit Istituto Nazionale di Fisica Nucleare, University and Scuola
        Normale Superiore of Pisa, I-56100 Pisa, Italy} \\
\r {23} {\eightit University of Pittsburgh, Pittsburgh, Pennsylvania 15260} \\
\r {24} {\eightit Purdue University, West Lafayette, Indiana 47907} \\
\r {25} {\eightit University of Rochester, Rochester, New York 14627} \\
\r {26} {\eightit Rockefeller University, New York, New York 10021} \\
\r {27} {\eightit Rutgers University, Piscataway, New Jersey 08854} \\
\r {28} {\eightit Academia Sinica, Taiwan 11529, Republic of China} \\
\r {29} {\eightit Superconducting Super Collider Laboratory, Dallas, 
Texas 75237} \\
\r {30} {\eightit Texas A\&M University, College Station, Texas 77843} \\
\r {31} {\eightit Texas Tech University, Lubbock, Texas 79409} \\
\r {32} {\eightit University of Tsukuba, Tsukuba, Ibaraki 305, Japan} \\
\r {33} {\eightit Tufts University, Medford, Massachusetts 02155} \\
\r {34} {\eightit University of Wisconsin, Madison, Wisconsin 53706} \\
\r {35} {\eightit Yale University, New Haven, Connecticut 06511} \\
\end{center}
\vspace{0.5cm}

\vfill

\begin{abstract}
We present the first measurement of associated direct photon + muon production 
in hadronic collisions, from a sample of 1.8~TeV $p \bar p$ collisions
recorded with the Collider Detector at Fermilab.
Quantum chromodynamics (QCD) predicts that these events are primarily from
the Compton scattering process $cg\rightarrow c\gamma$, 
with the final state charm quark producing a muon. 
Hence this measurement is sensitive to the charm quark 
content of the proton.  The measured cross section of 
$29\pm 9 $ pb is compared to 
a leading-order QCD parton shower model as well as a 
next-to-leading-order QCD calculation.
\end{abstract}
\end{center}
\vfill
\newpage
\end{titlepage}

\begin{figure}[tbp]
\begin{center}
\begin{minipage}[h]{6.5in}
\epsfxsize=6.3in
\epsfbox[36 144 520 650]{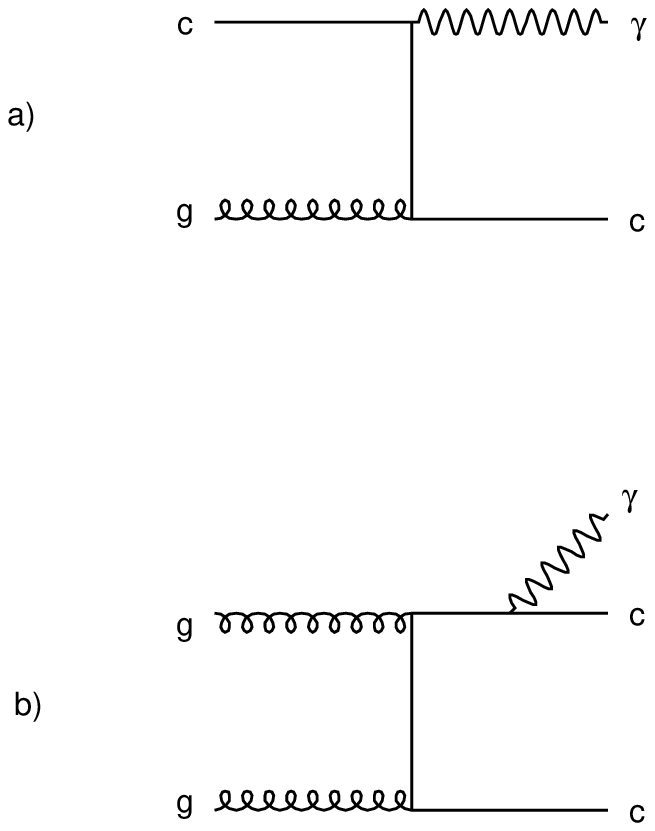}
\caption{Two diagrams depicting the QCD picture of 
photon+charm quark production.  In a) the dominant diagram is depicted, 
the Compton scattering process which enters at leading-order in QCD.
Fig. b) represents the largest process that enters the QCD calculation
for the first time at next-to-leading order.}
   \label{charmlonlo}
\end{minipage}
\end{center}
\end{figure}

Direct photon production in hadronic interactions has historically provided
precise tests of quantum chromodynamics (QCD). The associated production of
charm quarks in direct photon events provides a potential probe of the charm
quark content of the proton through the Compton scattering process $%
cg\rightarrow c\gamma $ as shown in Fig.~\ref{charmlonlo}a~\cite{nloqcd}. In
a previous publication~\cite{dstar} we tagged the charm quark with a full
reconstruction of a $D^{\ast }$ meson. In this Letter we present the first
measurement of the associated production cross section of $\gamma +\mu $ 
in hadron-hadron collisions. In addition, since the previous publication
next-to-leading order QCD (NLO QCD) calculations~\cite{nloqcd} have become
available.  At NLO some new processes, which produce a
significant increase in the number of photon+charm quark events, enter the
QCD calculations. The largest of these new
processes is $gg\rightarrow c\bar{c}\gamma $, which is depicted in Fig.~\ref
{charmlonlo}b.

The data for this analysis consists of an integrated luminosity of 13.2~pb$%
^{-1}$ of $p\bar{p}$ collisions collected by the Collider Detector at
Fermilab (CDF) in the 1992-93 Tevatron collider run (Run 1a). 
The CDF detector has
been described in detail elsewhere~\cite{det_cdf}.
The events in the 
photon data sample discussed in this paper contain a cluster of energy
in the central electromagnetic calorimeter $|\eta _{\gamma }|<0.9$, with no
charged tracks pointing to the cluster. The clusters are required to have a
transverse momentum $P_{T}(=Psin(\theta ))$ between 17 and 40~GeV and to be
isolated, with less than 2 GeV of additional transverse energy in a cone of $%
\Delta R=\sqrt{\Delta \phi ^{2}+\Delta \eta ^{2}}=0.7$ around the cluster.
Additional photon cuts were used which were identical to those used in the
Run 1a CDF inclusive photon 
analysis~\cite{cdfpho}, after these cuts 108K photon candidate events remain. 
Direct photon backgrounds
from $\pi ^{0}$ and $\eta $ meson decays remain in the sample. They are
subtracted on a statistical basis by using the photon background subtraction
``profile'' method described in reference~\cite{cdfpho}. 
The upper cut on photon $P_T$ described above is necessary in order
to use this technique.
Using this method, the data sample is estimated to contain 
$35\pm 7\%$ background from meson decays, leaving 70K direct photon events
before muon selection.  

\begin{figure}[tbp]
\begin{center}
\begin{minipage}[h]{6.5in}
\epsfxsize=6.3in
\epsfbox[36 144 520 650]{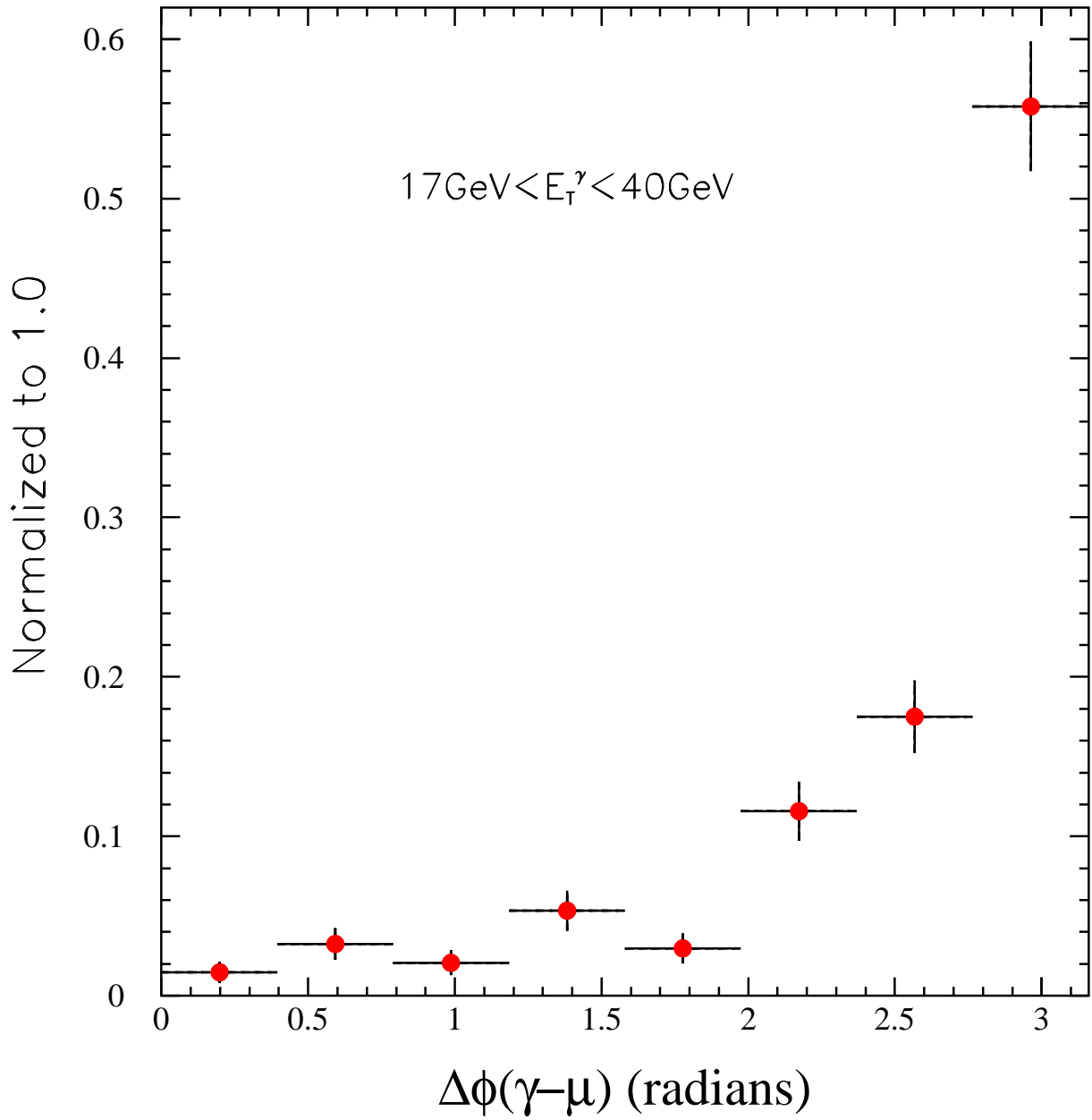}
	\caption{The difference in azimuthal angle between the photon and 
	the muon candidate is shown for the direct photon events.}
	\label{dphi}
\end{minipage}
\end{center}
\end{figure}

Muon candidates in this direct photon sample were selected by additionally
requiring a match between a charged track with $P_{T}>4$ GeV in the central
tracking chamber and a track in the appropriate muon system. For $|\eta
_{\mu }|<0.6$ the central muon system and central muon upgrade provide
mostly redundant coverage and matching tracks were required in each system,
while for $0.6<|\eta _{\mu }|<1.0$ the track was required to be in the
central muon extension. All three muon systems and their respective muon
identification are discussed in detail in reference~\cite{thesis}, the 
identification is similar to that used in reference~\cite{top}. After the
track matching requirement there are $140\pm 39$ direct photon events with
muon candidates (after the photon background subtraction described above).
The azimuthal angular difference between the direct photon and the muon
candidate is shown in Fig.~\ref{dphi}, showing that they are very
back-to-back as expected for the $2\rightarrow 2$ scattering process.

A number of quantities that define the quality of the muon candidates, such
as the position match between the track and the muon stub, were compared to
a clean sample of muons from J/$\psi $ decays, and were found to be in
excellent agreement. A small background due to accidental track-stub matches
was found in the central muon extension, and subtracted as discussed in
reference~\cite{thesis}. No corresponding background was found in the other
muon systems.
Note that some of the muon candidates are muons from
charged pion and kaon decays (decay-in-flight), while a smaller fraction are
from charged hadrons that do not interact significantly in the material in
front of the muon detectors (punch-through). 

\begin{figure}[tbp]
\begin{center}
\begin{minipage}[h]{6.5in}
\epsfxsize=6.3in
\epsfbox[36 144 520 650]{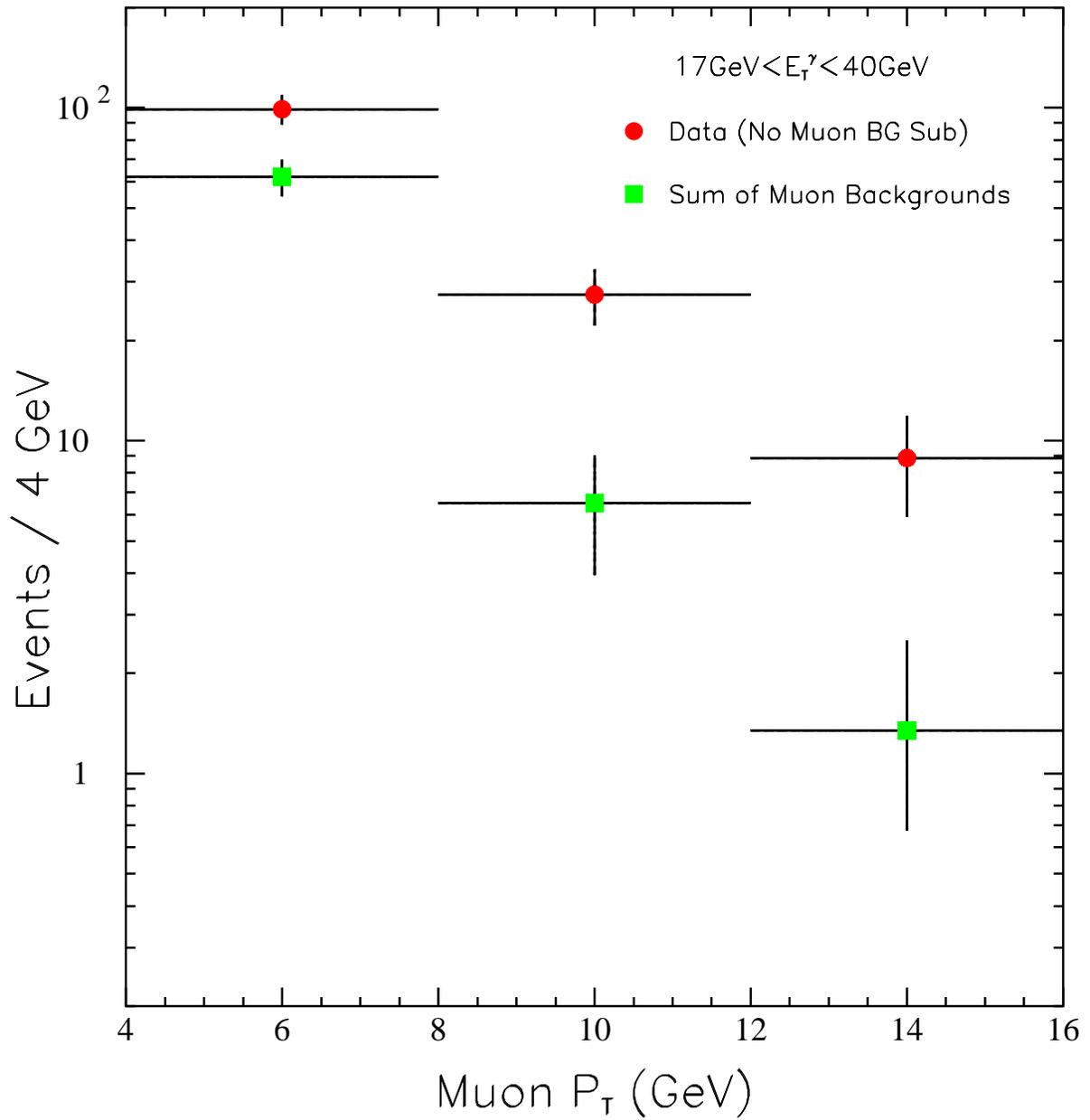}
	\caption{The muon pt distribution in direct photon events is shown.
	The sum of all muon backgrounds is also shown.}
	\label{ptmu}
\end{minipage}
\end{center}
\end{figure}

These backgrounds are estimated with a study that
includes both data and Monte Carlo simulation.  Starting with
the parent inclusive photon + jet data sample, where the main
muon backgrounds come from,
we measure the 4-vector of each charged
particle.  These charged particles include those from the
jet as well as from the underlying event in the collision.  We then
pass this 4-vector into a CDF detector simulation as a charged pion
or kaon or proton.  Finally the results of the simulation are passed
through the muon reconstruction,  and the number of charged hadrons
that faked a muon is recorded.  Approximately 75\% of the background
is from decay-in-flight.  This background estimate should be robust
since the charged particle spectrum is measured directly from
the parent data sample and the kinematics of decay-in-flight are
well known.
The backgrounds from non-interacting protons are negligible due to their 
relatively large interaction cross section, thus only the
particle fractions of pions and kaons need to be determined to complete the
muon background estimate. A number of pion and kaon particle fraction
measurements exist at LEP and the Tevatron~\cite{thesis}, and they are quite
consistent with one another. We have used the LEP measurements of 65\%
charged pion and 25\% charged kaon as the nominal estimate since the
kinematics of those measurements are best matched to the current analysis.
We have used the other measurements to determine the systematic uncertainty
(the largest deviation being 60\% pion and 20\% kaon). Combining the
simulation results with the particle fractions gives the muon background
rate in the direct photon sample. Fig.~\ref{ptmu} shows the muon $P_{T}$ in
the direct photon sample, along with the estimated muon backgrounds. One
sees that the number of muon candidates is significantly larger than the
estimated background. The estimated background can also be checked directly
by the data by studying pions from reconstructed $K_{s}$ decays. This
estimate agrees well with the simulation result, but is statistically
weaker, thus the simulation result will be used.

Subtracting the estimated background leaves $72 \pm 20$(statistical) direct
photon + muon signal events. The photon-muon cross section is derived from
the 72 photon-muon events by dividing by the luminosity, 13.2 $pb^{-1}$, and
the efficiencies for detecting the photon within $|\eta_{\gamma}|<0.9$ and
the muon within $|\eta_\mu|<1.0$. These efficiencies are measured by a
combination of simulation and data studies, which are described in the
earlier references. The photon efficiency is $\approx 33\%$ with a small
photon $P_T$ dependence, while the muon efficiency is $\approx 57\%$ and
depends slightly on the specific muon detector. The resulting photon-muon
cross section is $\sigma^{data} = 29 $~pb~$\pm 8$~pb~(statistical).

There are four significant systematic uncertainties on direct photon + muon
cross section: 1) 10\% from the muon background subtraction, which mostly
comes from the uncertainties in the estimated pion and kaon particle
fractions, 2) 9\% from the photon background subtraction uncertainty, 3) 5\%
from the uncertainty in the photon and muon cut efficiencies, and 4) 3.6\%
from the uncertainty in the CDF luminosity measurement. These added in
quadrature are a 4 pb uncertainty, much smaller than the statistical
uncertainty in the measurement. The measured photon-muon cross section
including systematic uncertainties is therefore $\sigma^{data} = 29 $~pb~$%
\pm 9$~pb~(statistical+systematic).

\begin{figure}[tbp]
\begin{center}
\begin{minipage}[h]{6.5in}
\epsfxsize=6.3in
\epsfbox[36 144 520 650]{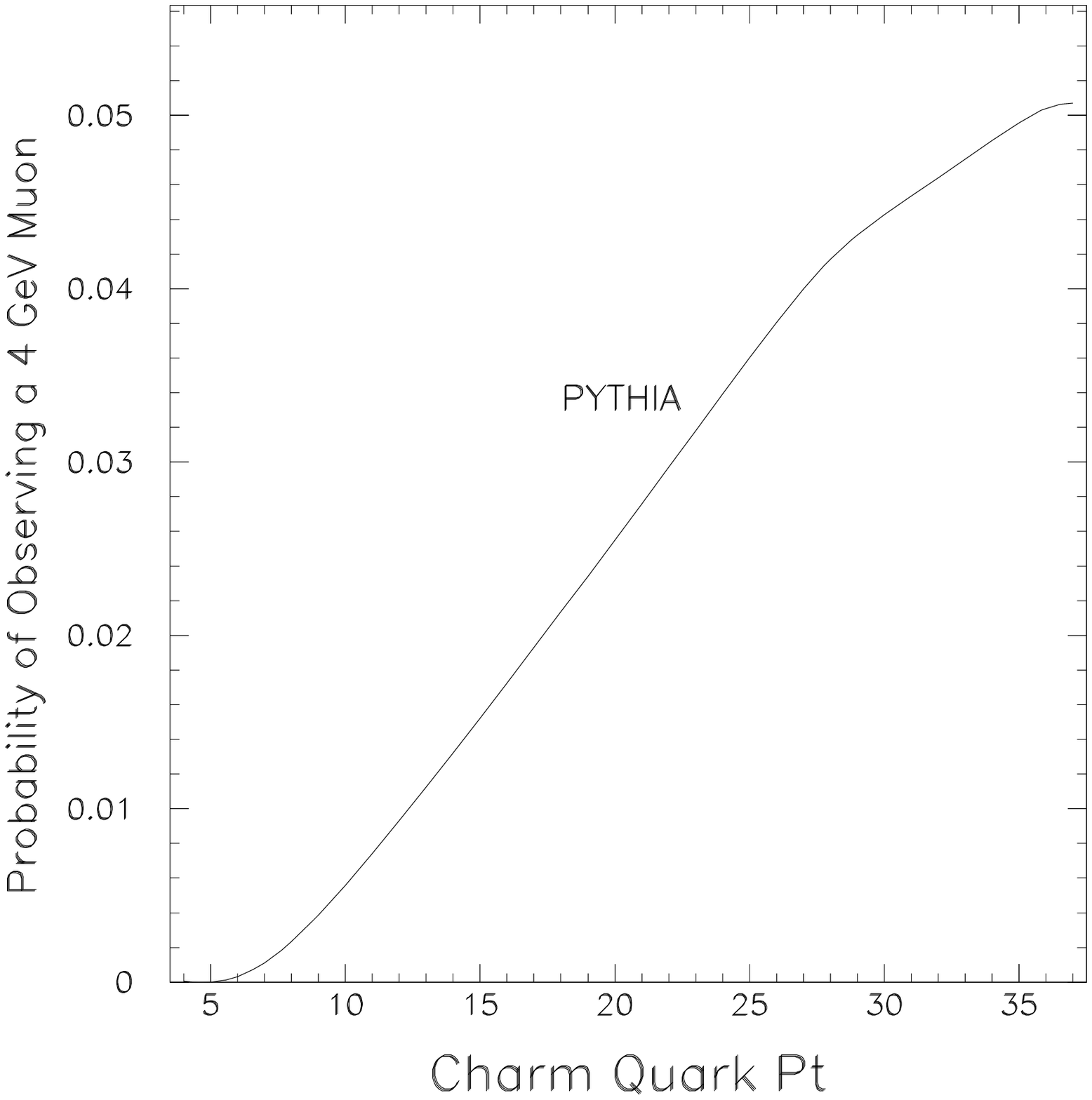}
	\caption{The probability from PYTHIA
	 for a charm quark to give a $>$4 GeV muon 
        is shown as a function of the charm quark $P_T$.}
	\label{turnon}
\end{minipage}
\end{center}
\end{figure}

The photon-muon cross section will be compared to two different QCD
calculations of photon-muon production.
The first QCD calculation is
the PYTHIA~\cite{pythia} Monte Carlo which only has the leading-order
contributions to the photon+heavy quark cross section, but has the full
parton shower and fragmentation effects, and includes the photon+bottom 
process (about 25\% of the total cross section).  The second QCD calculation 
is the NLO QCD photon+charm calculation detailed in 
reference~\cite{nloqcd}. This has 
additional processes not present at leading-order, of which 
Fig.~\ref{charmlonlo}b is the largest contribution. But with only the 
charm quark in the final state
in the NLO QCD calculation, we need to convolute it with the effects of
charm quark fragmentation. We will use the Peterson fragmentation model in
PYTHIA to do this. An example of the resulting probability of observing a 4
GeV muon from this model is shown in Fig.~\ref{turnon} as a function of
charm quark $P_T$.  This curve includes the 10\% branching ratio and 
the charm fragmentation effects.  This calculation uses the massless 
quark approximation, which is adequate for photon+charm since the 
scale of the process is well above the charm mass,  but cannot 
be applied to the photon+bottom process.

The resulting QCD cross sections for photon+muon production are tabulated in
table 1, along with the measured value.
While both models are consistent with the data
within two standard deviations, the NLO QCD predictions are significantly
larger than PYTHIA, and closer to the measured value. The ratio of the
measurement to PYTHIA is very similar to that observed in our previous
publication using reconstructed $D^{\ast }$s to tag the charm quark~\cite
{dstar}. A number of variations of the NLO QCD calculation were
investigated, such as renormalization scale, parton distributions, Peterson
parameters. The largest variations came from the Peterson parameters: a $\pm
0.02$ change in the fragmentation parameter caused the NLO QCD cross section
to change by $\pm 6\%$.

\begin{table}[htbp]
\begin{center}
\begin{tabular}{|c|c|c|}
\hline
Photon+Muon Data & PYTHIA & NLO QCD \\ 
& (CTEQ4L, $\mu=P_T/2$) & (CTEQ4M, $\mu=P_T/2$) \\ \hline
29 $\pm$ 9 pb & 13 pb & 21 pb \\ \hline
\end{tabular}
\end{center}
\caption{Comparison of the measured photon+muon cross section with the QCD
models.}
\end{table}

In summary, the first measurement of direct photon plus associated muon
production in hadronic interactions has been presented. Comparisons 
with NLO QCD photon+heavy~quark production have also been presented, and are
consistent with the measurement.

\newpage 
\clearpage

\clearpage

\clearpage

\clearpage


\begin{thebibliography}{9}
\bibitem{det_cdf}  F. Abe, {\em et al.}, Nucl. Inst. and Meth. {\bf A 271},
387 (1988), and the references therein. The coordinate system used within
the CDF is $(\theta,\phi)$, where $\theta$ is the polar angle relative to
the proton beam as measured from the event vertex, and $\phi$ the azimuth.
The pseudorapidity is defined as $\eta = -\ln \tan(\theta/2)$.
\end{thebibliography}
\end{document}